\documentclass[twocolumn,prd,showpacs,floatfix,preprintnumbers,nofootinbib,a4paper]{revtex4}

\usepackage{epsfig}

\newcommand{\gag}{g_{a\gamma}}
\def\lsim{\mathrel{\raise.3ex\hbox{$<$\kern-.75em\lower1ex\hbox{$\sim$}}}}
\def\gsim{\mathrel{\raise.3ex\hbox{$>$\kern-.75em\lower1ex\hbox{$\sim$}}}}

\begin{document}

\title{Detecting Axion-Like Particles With Gamma Ray Telescopes}

\author{Dan Hooper and Pasquale D.~Serpico}
\affiliation{Center for Particle Astrophysics,
Fermi National Accelerator Laboratory,
Batavia, IL 60510-0500 USA}

\date{\today}

\preprint{FERMILAB-PUB-07-190-A}

%%%%%%%%%%%%%%%%%%%%%%%%%%%%%%%%%%%%%%%%%%%%%%%%%%%%%%%%%%%%%%%%%%%%%%
\begin{abstract}
%%%%%%%%%%%%%%%%%%%%%%%%%%%%%%%%%%%%%%%%%%%%%%%%%%%%%%%%%%%%%%%%%%%%%%
We propose that axion-like particles (ALPs) with a two-photon
vertex, consistent with all astrophysical and laboratory bounds, may
lead to a detectable signature in the spectra of high-energy gamma
ray sources. This occurs as a result of gamma rays being converted
into ALPs in the magnetic fields of efficient astrophysical
accelerators according to the ``Hillas criterion", such as jets of
active galactic nuclei or hot spots of radio galaxies. The discovery
of such an effect is possible by GLAST in the 1-100 GeV range and by
ground based gamma ray telescopes in the TeV range.
\end{abstract}
%%%%%%%%%%%%%%%%%%%%%%%%%%%%%%%%%%%%%%%%%%%%%%%%%%%%%%%%%%%%%%%%%%%%%%

\maketitle

%%%%%%%%%%%%%%%%%%%%%%%%%%%%%%%%%%%%%%%%%%%%%%%%%%%%%%%%%%%%%%%%%%%%%%
{\it Introduction---}
%%%%%%%%%%%%%%%%%%%%%%%%%%%%%%%%%%%%%%%%%%%%%%%%%%%%%%%%%%%%%%%%%%%%%%
The Peccei-Quinn (PQ) mechanism~\cite{Pec77}
remains perhaps the most compelling explanation of the CP problem of
QCD. A new chiral $U_{\rm PQ}(1)$ symmetry that is spontaneously
broken at some large energy scale, $f_{\rm a}$, would allow for the
dynamical restoration of the CP symmetry in  strong interactions. An
inevitable consequence of this mechanism is the existence of axions,
the Nambu-Goldstone bosons of $U(1)_{\rm
PQ}$~\cite{Weinberg77}. One of the most
important phenomenological properties of the hypothetical axion is its two-photon
vertex which allows for axion-photon conversions in the presence of external electric
or magnetic fields~\cite{Dicus:1978fp} through an interaction
term
%%%%%%%%%%%%%%%%%%%%%%%%%%%%%%%%%%%%%%%%%%%%%%%%%%
\begin{equation}
     {\cal L}_{{\rm a}\gamma}=
     -\frac{1}{4}\,g_{{\rm a}\gamma} F_{\mu\nu}\tilde F^{\mu\nu}a
     =g_{{\rm a}\gamma}\,{\bf E}\cdot{\bf B}\,a\,,
   \label{eq1}
\end{equation}
%%%%%%%%%%%%%%%%%%%%%%%%%%%%%%%%%%%%%%%%%%%%%%%%%%
where $a$ is the axion field, $F$ is the electromagnetic field-strength
tensor, $\tilde F$ its dual, ${\bf E}$ the electric field, and ${\bf B}$
 the magnetic field. The axion-photon coupling strength is quantified
by
%%%%%%%%%%%%%%%%%%%%%%%%%%%%%%%%%%%%%%%%%%%%%%%%%%
\begin{equation}\label{eq2}
   g_{{\rm a}\gamma} =\xi
  \frac{\alpha}{2 \pi}\, \frac{1}{f_{\rm a}}\,
 \,,
\end{equation}
%%%%%%%%%%%%%%%%%%%%%%%%%%%%%%%%%%%%%%%%%%%%%%%%%%
where $\alpha$ is the fine-structure constant and $\xi$ is a
parameter of ${\cal O}$(1) depending on the details of the electromagnetic
and color anomalies of the axial current associated with the axion
field. In particular, this coupling is used by the ADMX experiment
to search for axion dark matter~\cite{Bradley:2003kg} and by the
CAST experiment to search for solar axions~\cite{Zioutas:2004hi,
Andriamonje:2007ew}. The Peccei-Quinn axion has the important
feature that its mass $m_a$ and interaction strength are inversely
related to each other and are connected to the measured properties
of pions. One may, however, conceive of a more general class of
particles whose coupling and mass are unrelated to each other. Such
states are known as axion-like particles (ALPs). ALPs may manifest
themselves in the propagation of photons in magnetic fields, either
in laboratory or astrophysical environments, and may have
potentially interesting astrophysical and cosmological
consequences~\cite{acconseq}.

In this letter, we propose another way to potentially detect ALPs,
namely through their distortion of the energy spectra of high-energy
gamma ray sources (we note however that a light scalar particle
coupling to $F_{\mu\nu}F^{\mu\nu}$ in Eq. (\ref{eq1}) would lead to
similar effects). This idea is somewhat similar to that discussed in
the recent Ref.~\cite{Mirizzi:2007hr}, but with some important
differences. In that paper the authors considered the ALP parameters
needed to fit PVLAS data~\cite{Zavattini:2005tm} (as in other
recently proposed gamma ray signatures of ALPs \cite{otherPVLAS}),
and assumed that the conversion of photons above $\sim 10$ TeV into
ALPs takes place in the turbulent component of the galactic magnetic
field. Here, in contrast, we discuss the case in which the
photon-ALP conversion occurs near or within the gamma ray sources.
Interestingly, we find that, if the gamma sources are (or are hosted
in) efficient astrophysical accelerators according to the ``Hillas
criterion" \cite{Hil84}, significant conversion can occur in ALP
models which are fully consistent with all laboratory and
astrophysical constraints. In fact, the mechanism discussed here may
offer the most practical way to detect ALPs over a significant range
of masses and couplings.

%%%%%%%%%%%%%%%%%%%%%%%%%%%%%%%%%%%%%%%%%%%%%%%%%%%%%%%%%%%%%%%%%%%%%%
{\it Photon-ALP conversion in gamma ray sources---}
%%%%%%%%%%%%%%%%%%%%%%%%%%%%%%%%%%%%%%%%%%%%%%%%%%%%%%%%%%%%%%%%%%%%%%
As a consequence of the interaction of Eq.~(\ref{eq1}), ALPs and
photons oscillate into each other in the presence of an external
magnetic field. For a photon of energy $E_{\gamma}$, the probability
of converting into an ALP can be written~\cite{mixingAG}
\begin{equation}\label{pertpz1a}
 P_{\rm osc}=\sin^2(2\theta)
 \sin^2\left[\frac{\gag\,B\,s}{2}\sqrt{1+\left(\frac{\mathcal{E}}{E_\gamma}\right)^2}\right]\,,
\end{equation}
where $s$ is the size of the domain and $B$ is the magnetic field
component along the polarization vector of the photon, which is
assumed to be approximately constant within that domain. We have
also defined an effective mixing angle $\theta$  and characteristic
energy $\mathcal{E}$ via
\begin{equation}\label{mixangE}
\sin^2(2\theta)=\frac{1}{1+(\mathcal{E}/E_{\gamma})^2}\,,\:\:\:\:\:\:\mathcal{E}\equiv\frac{m^2}{2\gag\,B}\,,
\end{equation}
where the effective ALP mass squared is $m^2\equiv
|m^2_{a}-\omega_{\rm pl}^2|$, $\omega_{\rm pl} =\sqrt{
4\pi\alpha\,n_e/m_e}$ is the plasma frequency,  $m_e$ the electron
mass, and $n_e$ the electron density. For the following
considerations, it is useful to introduce the dimensionless
quantities: $g_{11}=\gag/10^{-11}~{\rm GeV}^{-1}$, $B_{\rm G}\equiv
B/{\rm Gauss}$, $s_{\rm pc}\equiv s/{\rm parsec}$, $m_{\mu{\rm
eV}}\equiv m/\mu$eV, $\mathcal{E}_{\rm GeV}\equiv \mathcal{E}/$GeV.
Recent results from the CAST experiment~\cite{Andriamonje:2007ew}
provide a direct bound on the ALP-photon coupling of $g_{11}
\lesssim 8.8$ for $m_a\alt 0.02\,$eV, nominally below the
long-standing globular cluster limit~\cite{acconseq}. Note that
\begin{equation}\label{oscunits} \omega_{\rm pl}=
0.37\times 10^{-4}\,\mu{\rm eV}\sqrt{n_e/{\rm cm}^{-3}}\,,
\end{equation}
which means that in the interstellar medium (ISM) of the Milky Way,
where $n_e\sim0.1\,{\rm cm}^{-3}$, the effective mass of the ALP
will not be smaller than $m_{\mu{\rm eV}}\sim 10^{-5}$,
independently of how small $m_a$ is. For ultra-light ALPs ($m_a\alt
10^{-11}\,$eV), the absence of gamma rays from SN~1987A yields a
stringent limit of $g_{11}\lesssim 1$~\cite{Brockway:1996yr} or even
$g_{11} \lesssim 0.3$~\cite{Grifols:1996id}. But for $10^{-11}\,{\rm
eV}\ll m_a\ll 10^{-2}\,$eV, the CAST bound is the most general and
stringent (bounds from ADMX, although stronger for some masses,
assume that the axion is the Galactic dark matter).

{\it General properties---} From Eqs. (\ref{pertpz1a},\ref{mixangE})
it follows that: (i) At energies below $\mathcal{E}$, the mixing is small and {\it a
fortiori} the conversion probability is small. Above this critical
energy, the mixing is large, and a significant depletion probability
might arise. In suitable units,
\begin{equation}\label{cond1}
\mathcal{E}_{ \rm{GeV}}\equiv\frac{m_{\mu {\rm eV}}^2}{0.4\,
g_{11}\,B_{\rm G}}\,.
\end{equation}
As we shall argue, when plugging in the previous formula typical
astrophysical and ALP parameters, this critical energy naturally
falls in the gamma ray energy range. Thus, the physics of light and
weakly coupled ALPs naturally points to {\it $\gamma$-rays} as the
most promising tool for discovery. (ii) A significant conversion into axions also requires that the argument
of the oscillatory function in Eq. (\ref{pertpz1a}) is not too
small, i.e.
\begin{equation}\label{cond2}
\gag\,B\,s/2 \agt 1\,,\:\:\:{\rm i.e.}\:\:\:15\,g_{11}\,B_{\rm
G}\,s_{\rm pc}\gsim 1\,.
\end{equation}
The condition in Eq. (\ref{cond2}) depends on the product $B\,s$,
which also determines the maximum energy $E_{\rm max}$ to which
sources can confine and thus accelerate ultra-high energy cosmic
rays (UHECRs).  This is known as the Hillas criterion~\cite{Hil84},
and for protons it writes
\begin{equation}\label{HillEq}
E_{\rm max}\simeq 9.3\times 10^{20}\,{\rm eV}\, B_{\rm G} \,
s_{\rm{pc}}\,.
\end{equation}
This connection between cosmic ray acceleration and Eq.~(\ref{cond2}) 
is important. Since UHECRs with energies of a few
times 10$^{20}\,$eV have been observed, environments where $B_{\rm
G} \, s_{\rm{pc}} \agt 0.3$ must exist in nature. This implies that
couplings as small as $g_{11} \simeq 0.2$ might be probed, almost
two orders of magnitude below present bounds.

In the following, we shall examine in greater detail what are the
signatures expected in $\gamma$-rays due to the ALP-photon
conversion mechanism, what are the most promising sources to look
at, and the perspectives for current instruments to probe the ALP
parameter space.

%%%%%%%%%%%%%%%%%%%%%%%%%%% FIGURE 1 %%%%%%%%%%%%%%%%%%%%%%%%%%%%%%%%%
\begin{figure}
\centering \epsfig{figure=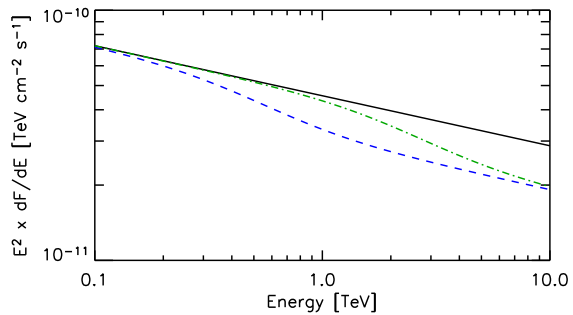,width=1.0\columnwidth,angle=0}
\caption{\label{fig1} A typical power-law $\gamma$-ray spectrum
(solid line) and its distortion for photon-ALP conversion with
$\mathcal{A}= 1/3$ and critical energies $\mathcal{E}=500\,$GeV
(dashed lines) and $\mathcal{E}=2.5\,$TeV (dashed-dotted line). See
text for details.}
\end{figure}
%%%%%%%%%%%%%%%%%%%%%%%%%%%%%%%%%%%%%%%%%%%%%%%%%%%%%%%%%%%%%%%%%%%%%%
{\it Spectral signatures---} The qualitative signatures of the
scenario considered here are remarkably robust, although the
quantitative aspects are model dependent. The reason for this is
twofold: (i) concerning particle physics, we ignore the fundamental
mass and coupling parameters $m_a$ and $\gag$; (ii) the complicated
(and unknown) 3-D field configurations typically present in
astrophysical environments do not allow to apply naively
Eq.~(\ref{pertpz1a}) for detailed quantitative predictions of the
magnitude of the depletion. Nonetheless, the feature one should look
for can be robustly parameterized as modification of the undistorted
spectrum $F_0(E_\gamma)$ into a modified spectrum
\begin{equation}\label{effective}
F_{a}(E_{\gamma})=\left[1-\frac{\mathcal{A}}{1+(\mathcal{E}/E_{\gamma})^2}\right]F_0(E_{\gamma})\,,
\end{equation}
where the constant $\mathcal{A}$ can be obtained in an idealized
case from the oscillatory sinus function in Eq. (\ref{pertpz1a}). As
an estimate, an efficient conversion of an unpolarized photon source
resulting in a complete depolarization of the photon-ALP system
would cause an average depletion of 33\% of the initial photon flux
(i.e. $\mathcal{A}\simeq 1/3$) above an energy $\mathcal{E}$
determined roughly by Eq.~(\ref{cond1}). In Fig. \ref{fig1} we plot
the spectral distortion for $\mathcal{A}= 1/3$ for the two cases
$\mathcal{E}=500\,$GeV and $\mathcal{E}=2.5\,$TeV, and for a
representative high energy gamma ray source. In particular, we
assumed $F_0(E_\gamma)=k\times E_\gamma^{-\Gamma}$, with
$\Gamma(=-2.2)$ and $k$ consistent around 100 GeV with the flux of
the blazar object Mkn 421 as reported by the MAGIC collaboration
\cite{Albert:2006jd}. This object is a powerful emitter also
observed by EGRET \cite{Hartman:1999fc} at GeV energy. Note that to
detect the signature expected, one needs a wide dynamical range and
sufficient statistics to detect a normalization shift of the typical
power-law spectra at the level of 10-20\%. Overall systematic errors
in the energy scale, in the aperture and in the exposure are
irrelevant to such a detection. Also, in variable sources the ALP
signature should be impressed over the variable spectrum at all
times, since it depends on the propagation and not on the emission
of the photons.) EGRET statistics and energy resolution are not
sufficient to look for such a signature, and only recently
ground-based gamma-ray telescopes have reached comparable
performances for powerful emitters within reasonable exposure times.
So, it would not be a surprise if such a signature had escaped
detection so far, but would show up in the coming years thanks to
the GLAST satellite detector and present and planned ground
telescopes.
In principle, if the astrophysical parameters were known, the
amplitude of the depletion could be used to constrain $\gag$, see
Eqs.~(\ref{pertpz1a},\ref{cond2}), while the energy at which the
effect is observed could be used to infer $m$, see
Eq.~(\ref{cond1}). If only upper limits were available for $B\,s$, a
lower limit on $\gag$ could be obtained, at least. Another
prediction is that if a hint for an ALP would show up in a source at
energy $\mathcal{E}_1$, then gamma emitters sitting in regions with
similar values of $B\,s$ should show a feature of similar amplitude
at a characteristic energy related to $\mathcal{E}_1$ only by the
value of their field strength, see Eq.~(\ref{cond1}).

%%%%%%%%%%%%%%%%%%%%%%%%%%% FIGURE 1 %%%%%%%%%%%%%%%%%%%%%%%%%%%%%%%%%
\begin{figure}
\centering \epsfig{figure=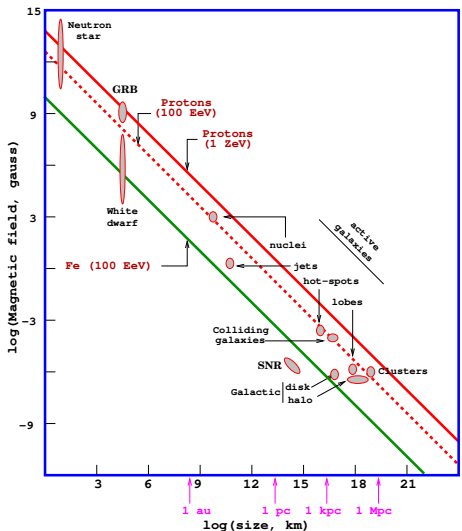,width=.7\columnwidth,angle=0}
\vspace{-1.0cm} \caption{Hillas diagram showing size and
magnetic field strengths of astrophysical objects required to
accelerate ultra-high energy cosmic rays (figure from
Ref.~\cite{Anchordoqui:2002hs} with permission). The Hillas condition is
closely related to the condition for the efficient conversion of
gamma rays into ALPs [see Eq.~(\ref{cond2})]. \label{fig2} }
\end{figure}
%%%%%%%%%%%%%%%%%%%%%%%%%%%%%%%%%%%%%%%%%%%%%%%%%%%%%%%%%%%%%%%%%%%%%%
{\it Promising sources---} To be consistent with existing bounds
\cite{Andriamonje:2007ew}, an ALP should have a coupling $g_{11}
\alt 9$, which implies that $B_{\rm G} \, s_{\rm{pc}} \agt 10^{-2}$
must hold at the source. In Fig.~\ref{fig2}, the Hillas diagram is
shown, reporting the typical $B$ and $s$ values for UHECR candidate
sources. It is clear that virtually  all the objects proposed as
UHECR accelerators, from gamma-ray bursts to clusters of galaxies,
appear suitable for the search of ALP signatures. This is fortunate,
since many observed (e.g. blazars) or expected (e.g. galaxy
clusters) gamma ray sources are hosted in or near putative UHECR
accelerators. However, at least the compact sources on the Hillas
plot are not likely the best candidates due to their higher
densities. Even a density of $10^{-6}$ g/cm$^3$, very low for
terrestrial standards, would imply $m\gg 1\,$meV [see
Eq.~(\ref{oscunits})] and thus the condition $E>\mathcal{E}$ cannot
be satisfied in the energy range probed by gamma ray astronomy. This
is a general, although qualitative, argument disfavoring too compact
(and presumably dense) sources as possible sites to observe
photon-ALP mixing. Very promising sources are instead AGN jets and
hot spots in radio galaxies such as Cygnus A and M87. For example,
typical parameters for the hot spots of Cygnus A are $B_{\rm
G}\simeq 0.15\times 10^{-3}$, $s_{\rm pc}\simeq 2\times
10^{3}$~\cite{Wilson:2000zu}, and similar numbers apply to the hot
spots of M87 \cite{Stawarz:2005wh}, which has been detected in the
TeV range. In these environments, the quantity of Eq.~(\ref{cond2})
is near unity for $g_{11}\sim 0.3$, while for propagation in our
galaxy the same coupling would fail to satisfy that condition by
more than one order of magnitude.

A remark is in order. If it were proved that conservative estimates
for the product $B\,s$ of a detected $\gamma$-ray emitter satisfy
Eq.~(\ref{cond2}) for $g_{11} \alt 9$, then gamma observations would
turn into powerful probes of ALP physics. But vice versa is not
necessarily true: indeed, although some fits assuming
synchrotron-self-Compton models seem to indicate that many detected
gamma ray sources (see e.g. Refs.~\cite{Albert:2006jd,othergamma})
fall short of the requirement of Eq.~(\ref{cond2}) by one order of
magnitude or more, it is important to remember that the ALP
conversion feature depends on the properties of the environment
crossed, not of the emitting region. In these cases, although a
negative result can not be used to put significant bounds, a
serendipitous discovery is by no means excluded.

\begin{figure}
\centering \epsfig{figure=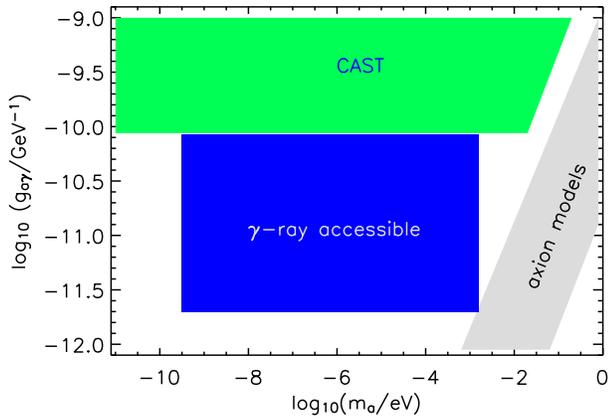,width=1.0\columnwidth,angle=0}
\caption{The approximate range of ALP parameters which could lead to
observable effects in gamma ray telescopes. Also shown are the
parameters excluded by CAST and the band preferred for QCD axion
models.\label{fig3} }
\end{figure}

{\it Exploring the ALP parameter space---} One can easily estimate
the range of ALP parameters observationally accessible. As we argued
earlier, from the highest energy UHECR observed the Hillas criterion
suggests {\it conservatively} that sites where $B_{\rm G} \,
s_{\rm{pc}} \agt 0.3$ must exist in nature. Once plugged into Eq.
(\ref{cond2}) (assuming equality), this implies that {\it at least}
couplings as small as $g_{11} \simeq 0.2$ may produce significant
depletions in gamma-ray spectra. To deduce  the range of masses
which can be probed, we proceed as follows: (i) we neglect too
compact objects on the Hillas plot and restrict our attention to the
most promising range of astrophysical source sizes previously
discussed, $s_{\rm pc}\sim 10^{-4}$ to $10^{5}$, deducing the
corresponding field strength (see Fig.~\ref{fig2}); (ii) we plug
these values in Eq. ~(\ref{cond1},) thus obtaining
${\mathcal E}_{ \rm{GeV}}\sim 4\times
10^{-3}\div 4\times 10^{6}\,m_{\mu {\rm eV}}^2\,.$
Unless $m \alt 0.1\mu$eV {\it and} the region is very compact (which
is disfavored, as previously discussed), the transition energy is
expected to fall in the gamma ray band, confirming again our
initial, general remarks. Considering the energy range most
interesting for GLAST (sub-GeV to a few tens of GeV), we deduce that
ALPs in the mass range $m_{\mu {\rm eV}}\sim{\rm few}\times
10^{-4}\div 10^{2}$ can be probed. For ground based gamma ray
telescopes such as HESS, MAGIC and VERITAS, which are sensitive to
gamma rays in the approximate range of $10^{2}$ to $10^4$ GeV, the
photon-ALP conversion can also be significant for masses in the
range of $m_{\mu {\rm eV}}\sim {\rm few}\times 10^{-3}$ to $10^{3}$.
Globally, we estimate the approximate range of parameters which
could lead to observable effects as the region schematically shown
in Fig.~\ref{fig3}, along with the range excluded by CAST and the
band preferred by QCD axion models~\cite{Andriamonje:2007ew}. The
particularly interesting region which overlaps with the QCD models
band corresponds to $\sim 10\,$TeV transition energies with typical
parameters of AGN cores.
%%%%%%%%%%%%%%%%%%%%%%%%%%% FIGURE 1 %%%%%%%%%%%%%%%%%%%%%%%%%%%%%%%%%
%%%%%%%%%%%%%%%%%%%%%%%%%%%%%%%%%%%%%%%%%%%%%%%%%%%%%%%%%%%%%%%%%%%%%%

{\it Summary---} Space and ground-based gamma ray telescopes may
have a chance to observe the effects of ALP-like particles (ALPs)
through the mechanism of photon-ALP mixing. This mechanism leads to
the depletion of gamma rays at high energies, resulting in a
peculiar signature in the spectra of gamma ray sources such as the
jets of active galactic nuclei, the hot spots of radio galaxies, or
clusters of galaxies. If the astrophysical parameters at the sources
were known sufficiently well, the mass and coupling of the ALP could
be reconstructed or, in case of negative outcome, excluded. Without
a detailed knowledge of the field strength and geometry around the
gamma source a detection is still possible, but a negative result
cannot easily be translated into interesting exclusion plots in the
ALP parameter space. We believe that the mechanism discussed here
should be thought of as an example of the opportunities that the new
generation of gamma ray telescopes will offer for studying
fundamental physics. Note that gamma ray telescopes may observe the
effects of ALPs with parameters which are exceedingly difficult to
explore otherwise, as shown in Fig.~\ref{fig3}. An appearance
experiment such as CAST is actually sensitive to $\gag^4$, and thus
would require an improvement of five to six orders of magnitude in
sensitivity to cover the entire range of parameters considered here.
However, a detection of the kind described here could be confirmed,
in part of the parameter space, by other astrophysical techniques,
such as that suggested in Ref.~\cite{Davoudiasl:2005nh}.

\smallskip
We would like to thank Tom Weiler for early discussions and comments
and G. Raffelt for comments on the manuscript. This work was
supported by the DOE and NASA grant NAG5-10842. Fermilab is
operated by Fermi Research Alliance, LLC under Contract No. DE-AC02-07CH11359 
with the US Department of Energy.

%%%%%%%%%%%%%%%%%%%%%%%%%%%%%%%%%%%%%%%%%%%%%%%%%%%%%%%%%%%%%%%%%%%%%%
%%%  Bibliography  %%%%%%%%%%%%%%%%%%%%%%%%%%%%%%%%%%%%%%%%%%%%%%%%%%%
%%%%%%%%%%%%%%%%%%%%%%%%%%%%%%%%%%%%%%%%%%%%%%%%%%%%%%%%%%%%%%%%%%%%%%

\end{document}